\documentclass[prl,reprint,superscriptaddress,noeprint]{revtex4-2}
\usepackage[T1]{fontenc}
\usepackage{physics}
\usepackage{amsmath,amsfonts,bbm, graphicx, color,natbib}
\usepackage{amssymb}
\usepackage{appendix}
\usepackage{relsize}
\usepackage{xcolor}
\usepackage{scalerel}
\usepackage{tikz}
\usepackage{pdfpages}
\usetikzlibrary{svg.path}

\definecolor{orcidlogocol}{HTML}{A6CE39}
\tikzset{
  orcidlogo/.pic={
    \fill[orcidlogocol] svg{M256,128c0,70.7-57.3,128-128,128C57.3,256,0,198.7,0,128C0,57.3,57.3,0,128,0C198.7,0,256,57.3,256,128z};
    \fill[white] svg{M86.3,186.2H70.9V79.1h15.4v48.4V186.2z}
                 svg{M108.9,79.1h41.6c39.6,0,57,28.3,57,53.6c0,27.5-21.5,53.6-56.8,53.6h-41.8V79.1z M124.3,172.4h24.5c34.9,0,42.9-26.5,42.9-39.7c0-21.5-13.7-39.7-43.7-39.7h-23.7V172.4z}
                 svg{M88.7,56.8c0,5.5-4.5,10.1-10.1,10.1c-5.6,0-10.1-4.6-10.1-10.1c0-5.6,4.5-10.1,10.1-10.1C84.2,46.7,88.7,51.3,88.7,56.8z};
  }
}

\newcommand\orcid[1]{\href{https://orcid.org/#1}{\mbox{\scalerel*{
\begin{tikzpicture}[yscale=-1,transform shape]
\pic{orcidlogo};
\end{tikzpicture}
}{|}}}}
\usepackage[colorlinks]{hyperref}

\hypersetup{
	bookmarksnumbered,
	pdfstartview={FitH},
	citecolor={darkgreen},
	linkcolor={darkblue},
	urlcolor={darkblue},
	pdfpagemode={UseOutlines}}
\definecolor{darkgreen}{RGB}{20,100,20}
\definecolor{darkblue}{RGB}{0,0,130}
\definecolor{darkred}{rgb}{.8,0,0}

\makeatletter
\AtBeginDocument{\let\LS@rot\@undefined}
\makeatother

\begin{document}

\title{Many-body molecule formation at a domain wall in a one-dimensional strongly interacting ultracold Fermi gas}

\author{Andrzej Syrwid\orcid{0000-0002-0973-4380}}
\affiliation{Department of Physics, The Royal Institute of Technology, Stockholm SE-10691, Sweden}

\author{Maciej \L{}ebek\orcid{0000-0003-4858-2460}}
\affiliation{Center for Theoretical Physics, Polish Academy of Sciences, Aleja Lotnik\'ow 32/46, 02-668 Warsaw, Poland}
\affiliation{Institute of Theoretical Physics, University of Warsaw, Pasteura 5, 02-093 Warsaw, Poland}

\author{Piotr T. Grochowski\orcid{0000-0002-9654-4824}}
\email{piotr@cft.edu.pl}
\affiliation{Center for Theoretical Physics, Polish Academy of Sciences, Aleja Lotnik\'ow 32/46, 02-668 Warsaw, Poland}
 \affiliation{ICFO - Institut de Ci\`encies Fot\`oniques, The Barcelona Institute of Science and Technology, Av. Carl Friedrich Gauss 3, 08860 Castelldefels (Barcelona), Spain}

\author{Kazimierz Rz\k{a}\.zewski\orcid{0000-0002-6082-3565}}
\affiliation{Center for Theoretical Physics, Polish Academy of Sciences, Aleja Lotnik\'ow 32/46, 02-668 Warsaw, Poland}
\date{\today}

\begin{abstract}
We analyze how the presence of the bound state on top of strong  intercomponent contact repulsion affects the dynamics of a two-component ultracold Fermi gas confined in a one-dimensional harmonic trap.
By performing full many-body numerical calculations, we retrieve dynamics of an initially phase separated state that has been utilized to excite the spin-dipole mode in experimental settings.
We observe an appearance of pairing correlations at the domain wall, heralding the onset of a molecular faction at the interlayer between the components.
We find that such a mechanism can be responsible for the stabilization of the phase separation.
\end{abstract}

\maketitle
\textit{Introduction}---For decades, investigations of multicomponent mixtures---from alloys and polymers to biological systems and glasses---have provided a deep insight into details of intercomponent interplay.
Specifically, an effective repulsion between constituents may result in their spatial separation~\cite{Pethick2008,Pitaevskii2016}
and induce an itinerant ferromagnetism in metals such as iron or nickel~\cite{Giorgini2008,Brando2016}, where electrons spontaneously form extended, spin-polarized domains.
This phenomenon can be understood within a simple mean-field framework proposed by Stoner, where a short-ranged screened Coulomb repulsion overcomes the Fermi pressure that favors a paramagnetic state~\cite{Stoner1933}.
While the simplified Stoner approach allows for qualitative description of many-electron systems, it does not capture effects related to beyond short-range interactions that may promote different, competing mechanisms suppressing ferromagnetism~\cite{Saxena2000,Pfleiderer2001}.

In this Letter we focus on the exemplary case of a two-component atomic Fermi gas with tunable short-range repulsive interactions, for which the stability of a ferromagnetic state has been debated both in theory~\cite{Sogo2002,Karpiuk2004,Duine2005,LeBlanc2009,Conduit2009,Cui2010,Pilati2010,Chang2011,Pekker2011,Massignan2011,Massignan2014,Levinsen2015,Trappe2016,Miyakawa2017,Koutentakis2019,Grochowski2017a,Ryszkiewicz2020,Karpiuk2020,Koutentakis2020} and in experiment~\cite{DeMarco2002,Du2008,Jo2009,Sommer2011,Sanner2012,Lee2012,Valtolina2016,Amico2018}.
It stems from the fact that a repulsive interaction potential due to the Feshbach resonance supports a weakly bound molecular state~\cite{Chin2010}.
Then, ferromagnetic correlations can only manifest themselves in an excited state of the many-body system in contrast to the superfluid ground state of paired atoms.

Since late 2000s experiments have tried to settle whether pairing processes prevent the ferromagnetic domains from appearing.
In the initial attempts, some signatures, such as increase of a kinetic energy, suggested the onset of ferromagnetism, however these efforts proved inconclusive~\cite{Jo2009,Sanner2012}.
Only after the system was initialized in an artificial domain structure---in which components reside in their respective halves of a harmonic trap---the phase separation undoubtedly persisted for some finite time in a strong interaction regime~\cite{Sommer2011,Valtolina2016}.
Recently, time-resolved investigation of the competition between pairing and ferromagnetic instabilities provided deeper insight into many-body physics governing the dynamics in a quenched system~\cite{Amico2018}.

On the theory side, approaches neglecting pairing processes showed that a ferromagnetic transition should take place in a three-dimensional geometry~\cite{Duine2005,Conduit2009,Pilati2010,Chang2011,Heiselberg2011,Recati2011,Goulko2011,Palestini2012,He2012,He2016}.
However, inclusion of pairing has been done only at the mean-field level or by the introduction of phenomenological terms.
In the context of Stoner ferromagnetism, an analysis of many-body eigenstates has been performed in few-body one-dimensional (1D) systems with contact interactions~\cite{Bugnion2013,Sowinski2013,Gharashi2013,Lindgren2014}.
Nevertheless, in such a case the pairing process is supported only by purely attractive interactions in the absence of the repulsive core characterizing realistic atomic potentials. 

\begin{figure}[t]
\includegraphics[width=1.0\linewidth]{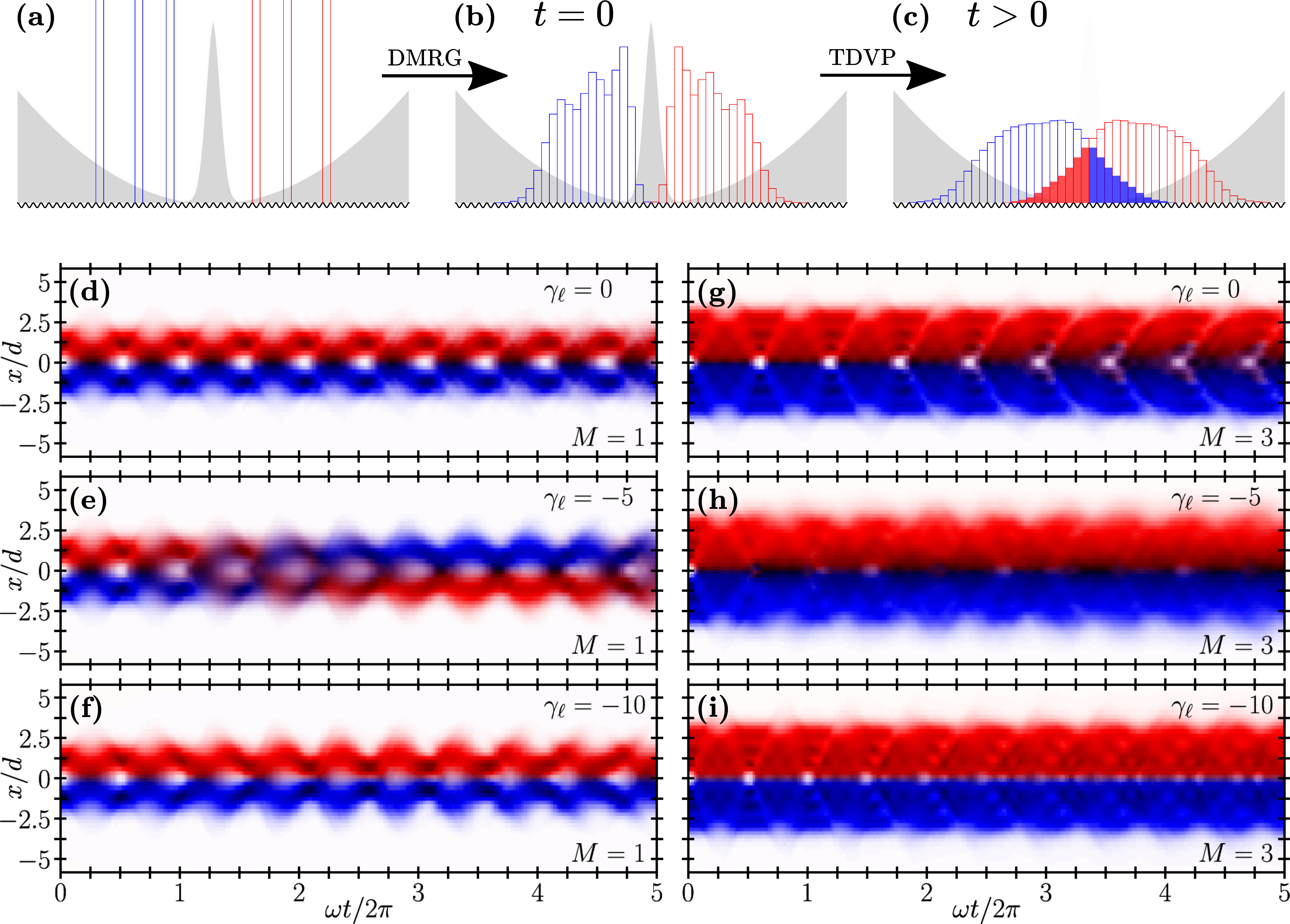}
	\caption{\label{fig1} 
	Illustration of how the phase separated state with $M=3$ is prepared (a)--(b) and then subsequently evolved after the removal of the central barrier (c).
	Time evolution of single-particle densities of two Fermi components interacting via the three-delta potential obtained for $M=1$, (d)--(f), and $M=3$, (g)--(i).
	For sufficiently weak and sufficiently strong beyond-contact attraction $\gamma_{\ell}$, the domain structure is stabilized for both $M=1$ and $M=3$.
	In contrast to the $M=1$ case, at the moderate attraction ($\gamma_{\ell} \approx \gamma_{\text{crit}}=-5$), $M=3$ system remains separated, suggesting an additional, many-body stabilization mechanism.
}
\end{figure}

To capture short-range details of the interatomic potential in 1D, we utilize the so-called three-delta potential~\cite{Gesztesy1985,Seba1986,Seba1986a,Cheon1998,Cheon1999,Veksler2014,Veksler2016,Jachymski2017}:
\begin{eqnarray} \label{3delta}
    W(x)= c_0 \delta (x)  + c_{\ell} \delta (x-\ell)  + c_{\ell} \delta (x+\ell), 
\end{eqnarray}
where $x$ denotes the relative position between fermions belonging to different spin components.
The parameters $c_0 >0$ and $c_{\ell} <0$ describe contact repulsion and finite-range attraction at a distance $\ell$, respectively.
It can be understood as a first correction to the contact repulsion due to finite-range interactions and was shown to reproduce van der Waals forces under quasi-1D confinement~\cite{Jachymski2017}.
Additionally, the intracomponent interactions are assumed to be negligible, as the system is brought close to the Feshbach resonance of the opposite spins and the temperature is very low.
The paradigmatic scenario considered here involves a strongly repulsive core with a weakly attractive well  ($0<-c_{\ell} \ll c_0$).
Our aim is to thoroughly analyze dynamics of two initially separated Fermi clouds---a setup inspired by former experiments~\cite{Valtolina2016,Peotta2012a,Ozaki2012}, see Fig.~\ref{fig1}(b). 
We argue that the potential $W(x)$ can give insight to many-body processes involved in the competition between pairing and ferromagnetic instabilities.

\textit{Model}---Mapping continuous, extended interactions onto discrete set of delta potentials was first analyzed in 1980s~\cite{Gesztesy1985,Seba1986,Seba1986a}.
It was soon realized that the three-delta potential is a minimal extension of a realistic short-range potential between atoms~\cite{Cheon1998,Cheon1999} and later utilized to study extended versions of the Gross-Pitaevskii and Lieb-Liniger models~\cite{Veksler2014,Veksler2016}.
Recently, it was shown that the three-delta potential provides valuable insight into qualitative behavior of the many-body system that goes beyond standard contact interactions~\cite{Jachymski2017}.

It should be stressed that in quasi-1D, the scattering theory implies a relationship $c_0+ 2 c_{\ell} = g_{\text{1D}}$, where $g_{\text{1D}}$ is a 1D mean-field coupling constant.
However, in our analysis these parameters are kept free, since we  consider full many-body model incorporating interplay between contact repulsion and beyond-contact attraction.
Note that such a model should also be attainable in experiments involving optical lattices, where on-site and nearest-neighbor interactions can be optically tuned.

For $\ell$ much smaller than other system length scales, it was shown that a simple contact interaction description can be retrieved with the effective coupling constant $c_{\mathrm{eff}}=g_\text{1D}$ + corrections depending on $\ell,c_0,c_{\ell}$ (for details see~\cite{Veksler2016}).
In our work, the length scale associated with $\ell$ is finite which affects the system beyond the contact approximation.

Our considerations are restricted to a balanced system consisting of $M$ identical fermions with mass $m$ in each of spin-$\uparrow$ and spin-$\downarrow$ components.
At the beginning we analyze a two-body case ($M=1$) determining a ground state wave function $\Psi(x_\uparrow,x_\downarrow)$
both for a free space and a harmonic potential confinement given by $\omega$ frequency, $U(x)=\frac{1}{2} m \omega^2 x^2$, for details see Supplemental Materials (SM).
While the point-like repulsion characterized by $c_0$ is assumed to be extremely strong, we modify the weak attraction strength $c_\ell$.
Consequently, $\Psi$ has a distinct cusp along $x_\uparrow=x_\downarrow$.
However, an increase of $|c_\ell|$ entails an enhancement of $|\Psi|^2$ around $|x_\uparrow -x_\downarrow|\approx\ell$, which is a premise of a bound state formation.
Indeed, by energy considerations we found that for each $c_0$ there exists some critical value of $c_\ell$ below which $\Psi$ represents a bound state (see SM).
As we show later, many-body dynamics of initially separated components with $M>1$ reveals a similar probability density accumulation for finding two fermions with opposite spins at a distance $\ell$, heralding the onset of the bound state. 

For large $M$ the many-body ground state can be studied within the lowest-order constrained variational (LOCV) approximation~\cite{Pandharipande1973,Pandharipande1977,Cowell2002,Heiselberg2011,Taylor2011,Yu2011,Grochowski2020b}, where the many-body wave function of the homogeneous, balanced two-component Fermi system is assumed to be represented by a product of two Slater determinants and a Jastrow factor, managing two-body correlations explicitly~\cite{Jastrow1955} (see SM).
Within the LOCV approximation, the Jastrow factor is varied to minimize the lowest order expanded total energy.
Such an approach has proved to be very accurate in the description of three-dimensional quantum mixtures, rivalling quantum Monte Carlo's accuracy.

Taking the limit $c_0\rightarrow \infty$ we found that the interaction energy in the ground state within the LOCV approximation becomes negative for $c_\ell \leq c_{\text{crit}}=-\frac{\hbar^2}{m\ell}$.
Note that the critical value $c_{\text{crit}}$ does not depend on $M$, corresponding to $c_{\text{eff}}=0$ from~\cite{Veksler2016} and agrees with a $M=1$ exact solution (see SM).
Therefore, when crossing $c_{\text{crit}}$, we expect to observe a change of dynamical behavior of the initially separated ferromagnetic state, as signatures of bound structures may appear.

We now proceed to study dynamical properties of the initially phase separated system, where the fermionic components are confined to their respective halves of the harmonic trap by means of an extremely high potential barrier in the center.
Both components do not have any meaningful overlap and are effectively noninteracting.
The barrier is then instantaneously released and the system evolves freely with two components starting to push into each other. 
As the intercomponent overlap starts to accumulate, the corresponding correlations begin to appear around the trap center.

\textit{Two-body dynamics}---First, we focus on $M=1$ case which can be analyzed analytically (see SM).
We stick to the case in which $\ell/d \approx  0.238$, where $d=\sqrt{\hbar / m \omega}$ is the harmonic trap length.
For convenience, the following dimensionless interaction parameters, $\gamma_\alpha=2^{1/4}\frac{c_\alpha}{d\hbar \omega}$ with $\alpha=0,\ell,\text{crit}$, are utilized.
Throughout our analysis, the repulsive core constant is set to $\gamma_0 = 100$ ($ \gamma_0\gg|\gamma_{\ell}|$), corresponding to large $g_{\text{1D}}$, which guarantees a phase separation of fermionic clouds at the mean-field level.

Despite being very weak in comparison to the zero-range repulsion, the beyond-contact attraction has dramatic consequences on the system dynamics.
When it is sufficiently weak, the stabilization of the initial phase separation is guaranteed  due to the central core repulsion, cf. Fig.~\ref{fig1}(d).
Surprisingly, the stabilization is also present when  the finite-range attraction is sufficiently strong, see Fig.~\ref{fig1}(f). 
Such a behavior can be tracked back to a system interacting via only contact potential undergoing an interaction quench from an infinitely repulsive Tonks-Girardeau (TG) gas~\cite{Tonks1936,Girardeau1960} to an excited super-Tonks-Girardeau (sTG) gaslike state of infinitely attractive atoms ~\cite{Astrakharchik2005,Syrwid2020}.
The initial state we consider can be decomposed into few lowest TG eigenstates due to the spatial separation between atoms, while $\gamma_{\ell} < \gamma_{\text{crit}}$ provides an effective strong attraction, despite large $g_\text{1D}$.
From the exact solution one can see, that in the limit $\gamma_{\ell} \to -\infty$ the dynamics of the system approaches the dynamics under Hamiltonian with the same $\ell$ but with $\gamma_{\ell} \to \infty$ (see SM).
Further analysis of this mechanism will be provided in the subsequent work~\cite{Lebek2021a}.
On the other hand, intermediate regime, $\gamma_{\ell} \approx \gamma_{\text{crit}}$, does not support stabilization of the spatial separation, as atoms mix with each other, cf. Fig.~\ref{fig1}(e).
Let us now proceed to a similar analysis in the few-body system.

\textit{Many-body dynamics}---In order to tackle the many-body problem, we study the Fermi-Hubbard model representing a discretized version of the continuous model discussed above.
The contact interaction becomes the on-site one, while peripheral deltas correspond to the nearest-neighbor attraction.
More specifically, we restrict ourselves to the lattice of $50$ sites, where the lattice spacing $\Delta$ equals the distance between deltas, $\Delta = \ell$.
To prepare the system in the initially separated state we employ a standard density matrix renormalization group (DMRG) method in the presence of the harmonic trap and a Gaussian barrier.
It yields a matrix product state (MPS) in which the clouds of atoms belonging to different components are well separated and they are mirror images of each other with respect to the trap center, cf. Fig.~\ref{fig1}(b).
After removing the barrier at time $t=0$ the system dynamics is investigated with the help of a recently developed algorithm combining one-site time-dependent variational principle (TDVP) procedure~\cite{Kramer2008} and a global basis expansion~\cite{tdvp2020}, cf. Fig.~\ref{fig1}(c).
For both DMRG and time evolution we use ITensor C++ library~\cite{itensor}, where the codes for modified TDVP we employed are available thanks to the authors of~\cite{tdvp2020} (for numerical details see SM).

\begin{figure}[ht]
	\includegraphics[width=1\linewidth]{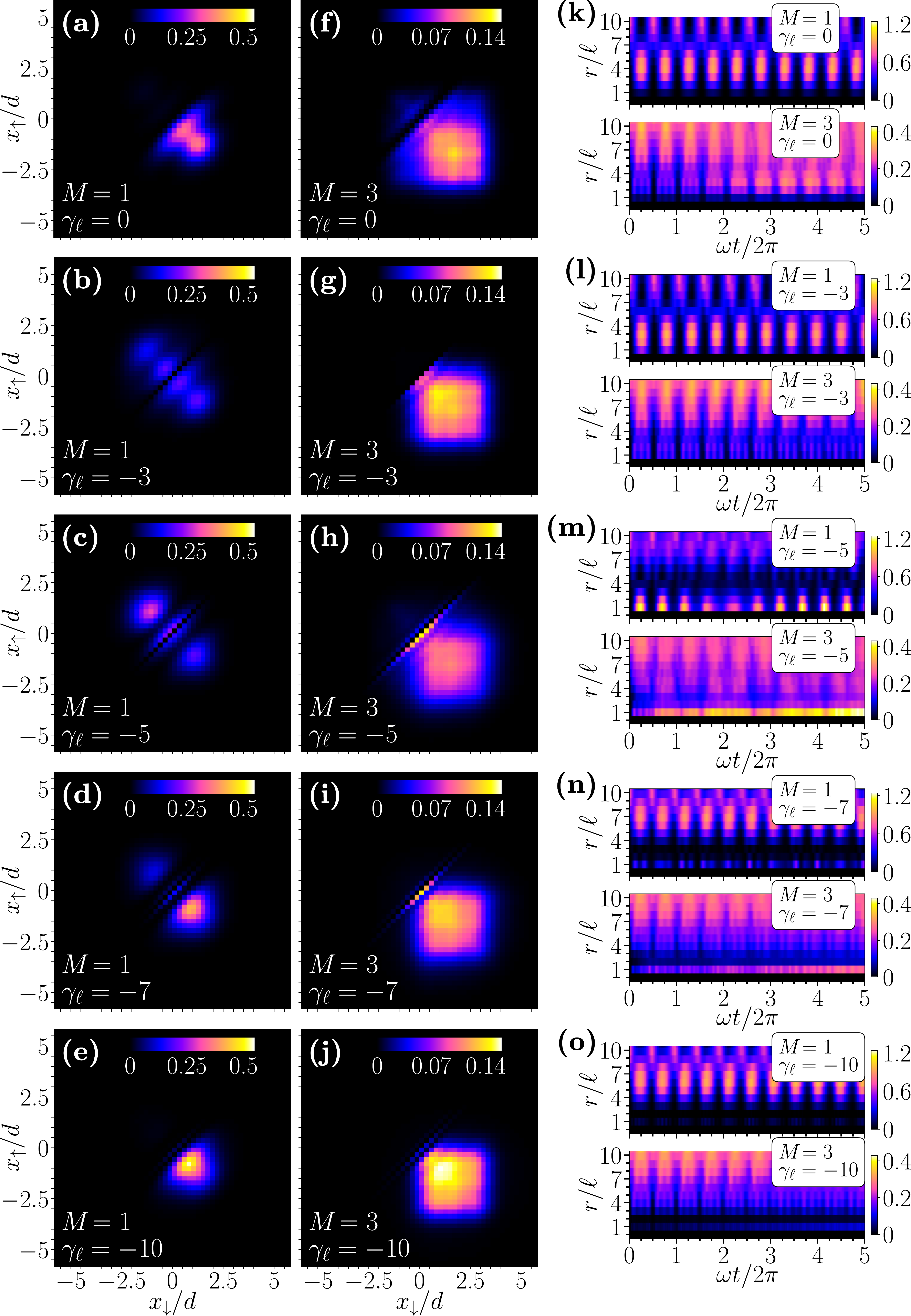}
	\caption{
	\label{fig2} 
	Temporally averaged over 5 trap periods density-density correlation functions $\langle G\rangle(x_\uparrow,x_\downarrow)$ calculated for $M=1$, panels (a)--(e), and $M=3$, panels (f)--(j), and for different finite-range attraction strengths $\gamma_\ell$.
	In accordance with the results shown in Fig.~\ref{fig1},  mixing between the two Fermi clouds takes place only for $M=1$ in the presence of intermediate attraction, see (b)--(d). 
	In contrast to the $M=1$ case, for $M=3$ with moderate attraction one can observe an accumulation of $\langle G\rangle(x_\uparrow,x_\downarrow)$ in the trap center, see (g)--(i). 
	In right panels (k)--(o) we show a temporal behavior of $g(r;t)$ obtained for $M=1,3$ and different $\gamma_\ell$. While for $M=1$ the pair distribution $g(r;t)$ oscillates together with single-particle densities plotted in Fig.~\ref{fig1}(d)--(f), for $\gamma_\ell \approx \gamma_\text{crit}$ the $M=3$ case reveals a steady and gradual increase of correlations between opposite-spin fermions separated by $r\approx\ell$. Such correlations are strictly related to the  presence of the finite-range attraction and are responsible for a domain stabilization.
	}
\end{figure}

In contrast to the $M=1$ case, the domain structure remains stable for all the interaction strengths (see Fig.~\ref{fig1}).
It can be explained by analyzing the intercomponent density-density correlation function $G(x,y;t)=\left<\Psi(t)\right|\hat{n}_\uparrow (x)  \hat{n}_\downarrow (y)  \left|\Psi(t)\right>/M^2$, with $\hat{n}_\sigma(x)=\hat{\psi}^\dagger_\sigma(x)\hat{\psi}_\sigma(x)$, where $\hat{\psi}_\sigma(x)$ is a canonical Fermi field operator for spin $\sigma$ at $x$ and $\left|\Psi(t)\right>$ is the time-evolved MPS at time $t$.
The correlation function is normalized such that $\int G(x,y;t)\mathrm{d}x \mathrm{d}y=1$.
In Fig.~\ref{fig2} we compare $\langle G\rangle(x_\uparrow,x_\downarrow)$ obtained for $M=1,3$ and different finite-range attraction strength $\gamma_\ell$, where $\langle...\rangle$ denotes a temporal average over 5 trap periods.
Far from $\gamma_\text{crit} $ both systems evolve similarly, suggesting that stabilization mechanisms described in the previous section are present also for $M>1$.
On the other hand, in the intermediate regime $\gamma_{\ell} \approx \gamma_{\text{crit}}$, an escalation of $\langle G\rangle$ at $|x_\uparrow -x_\downarrow| \approx \ell$ is revealed for $M=3$.
We interpret this enhancement as the onset of the bound state contribution (molecule faction) appearing in the course of time evolution.
It can be viewed as an extension of similar structures present in the relative wave function of the bound state in the two-body problem (see SM).
It is also supported by the fact that this growth takes place at the critical interaction strength predicted by the LOCV approximation and $M=1$ exact solution.
Note that in the $M=1$ case, the dynamics reveals no molecular formation as there are no additional atoms to absorb excess kinetic energy. 

This effect can be further investigated by analysis of the pair distribution function $g(r;t)=\int K_r(x;t) \mathrm{d}x$, where $K_r(x;t)=\frac{1}{2}\sum_{q=\pm}[G(x,x+q r;t)+G(x+q r,x;t)]$ describes spatial correlations between spin-$\uparrow$ and spin-$\downarrow$ fermions at a distance $r$.
The value of $g(r;t)$ corresponds to the probability density of finding two fermions of opposite spins at a distance $r$.
A temporal variability of $g(r;t)$ obtained for $M=1,3$ and different $\gamma_\ell$ is presented in Fig.~\ref{fig2}(k)--(o).
For $M=1$, independently of the $\gamma_\ell$ value, $g(r;t)$ oscillates, closely following the dynamics of single-particle densities, cf. Fig.~\ref{fig1}(e)--(f) and upper panels of Fig.~\ref{fig2}(k)--(o).
In contrast, for $\gamma_\ell\approx \gamma_\text{crit}$, the $M=3$ case reveals a steady and gradual growth of $g(r;t)$ at $r=\ell$, being the distance at which signatures of anticipated bound pairs are expected. 
Since the dominant contribution to $g(\ell;t)$ comes from the trap center, cf. panels (h) and (i) of Fig.~\ref{fig2}, we interpret this result as a footprint of molecular faction in the  interlayer between components.
Far from $\gamma_\ell\approx \gamma_\text{crit}$, the resulting $g(\ell;t)$ is inappreciable, 
suggesting that stabilization mechanisms noticed and described in the two-body case are also present in larger systems.

\begin{figure}[ht] 
	\includegraphics[width=1\linewidth]{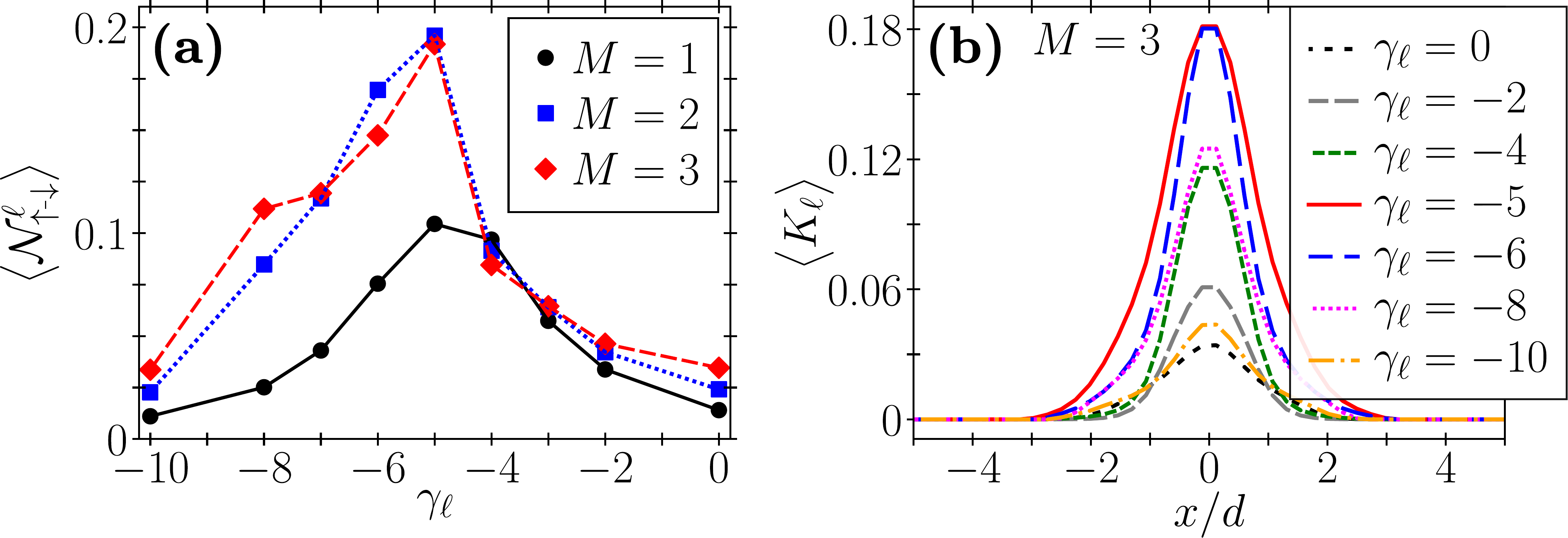}
	\caption{
	\label{fig3}
	Panel (a) shows the expectation value of the number of molecules $\langle \mathcal{N}_{\uparrow\text{-}\downarrow}^\ell \rangle$ averaged over 5 trap periods versus finite-range attraction strength $\gamma_\ell$.
	The results obtained for $M=2$ and $M=3$ are very similar and reveal a dramatic growth of $\uparrow$-$\downarrow$ pair correlations at a distance $r\approx\ell$ for $\gamma_\ell \approx-5$.
	On the other hand, for $M=1$, an increase of $\langle \mathcal{N}_{\uparrow\text{-}\downarrow}^\ell \rangle$ is much less pronounced, which may indicate lack of the molecule formation in the two-body system. 
	Panel (b) presents $M=3$ case of $\langle K_\ell \rangle(x)$ determined for different beyond-contact attraction strengths $\gamma_\ell$. The correlations are significantly amplified for $\gamma_\ell\approx \gamma_\text{crit}=-5$ and reveal maximum in the trap center, suggesting that the molecular faction accumulates between two Fermi clouds.
	}
\end{figure}

To study the molecular faction in more detail, one can compute $\mathcal{N}_{\uparrow\text{-}\downarrow}^\ell=M g(\ell;t) \ell$ representing the expectation value of the number of $\uparrow$-$\downarrow$ pairs of size $\approx\ell$.
Since the components in the initial state are spatially separated, the anticipated $\uparrow$-$\downarrow$ bound structures may appear in the course of time evolution.
Therefore, in Fig.~\ref{fig3}(a) we investigate how the temporal average $\langle \mathcal{N}_{\uparrow\text{-}\downarrow}^\ell \rangle $ depends on $M$ and the beyond-contact attraction $\gamma_\ell$.
It is striking that while $\langle \mathcal{N}_{\uparrow\text{-}\downarrow}^\ell \rangle $ turns out to be almost $M$-independent for $\gamma_\ell>\gamma_\text{crit}$, at $\gamma_\ell\approx \gamma_\text{crit}$ one observes an abrupt splitting between the results obtained for $M=1$ and $M >1$.
For both $M=2$ and $M=3$, $\langle \mathcal{N}_{\uparrow\text{-}\downarrow}^\ell \rangle $ is very similar and reveals a dramatic growth up to $\approx 0.2$.
When $\gamma_\ell$ becomes more negative, it rapidly decays approaching the values obtained for $M=1$.  
The escalation of  $\langle \mathcal{N}_{\uparrow\text{-}\downarrow}^\ell \rangle $ for $M>1$ around $\gamma_\text{crit}$
can be better understood when looking at the spatial distribution  $\langle K_\ell \rangle (x)$.
That is, as shown for $M=3$ in Fig.~\ref{fig3}(b), in the vicinity of $\gamma_\text{crit}$ there is a significant accumulation of $\uparrow$-$\downarrow$ pairs of size $\ell$ in the center of the trap.
This is an additional signature of an appearance of a molecular faction that resides between the two components forming a domain wall and thus providing a phase separation mechanism.
It is reminiscent of the microemulsion of two fermionic components and dimers investigated in Refs.~\cite{Amico2018,Scazza2020}, where in the three-dimensional trap bound pairs were also present at the interlayer separating components.

\textit{Short-range atomic potentials}---Despite the fact that the model Hamiltonian we consider could be analyzed experimentally through geometries involving optical lattices, the paradigmatic case of the Stoner instability in a two-component Fermi gas not constrained to the additional lattice is of particular interest.
In quasi-1D settings the perpendicular confinement may be very strong and confining length scales become comparable to the range of beyond-contact interactions making effects of finite-range coupling vastly pronounced~\cite{Jachymski2017}.  

In such  settings, finite-range corrections to the usual contact interaction can be modelled by the pseudopotential $W_p(r) = g_{\text{1D}} (1 + g' p^2) \delta(r)$, where $p = - i \dd / \dd r$ is the 1D momentum operator and $g_{\text{1D}}$ and $g'$ can be determined in a realistic scenario involving the Feshbach resonance and strong perpendicular confinement  (see Ref.~\cite{Jachymski2017} and SM for details).
When $p$ is discretized, the three-delta potential~\eqref{3delta} is recovered, however there is some ambiguity---there are multiple choices of $c_0$, $c_{\ell}$ and $\ell$ that correspond to the same parameters $g_{\text{1D}}$ and $g'$ describing identical scattering properties (see SM for a detailed calculation).
To map the realistic potential onto the three-delta one unambiguously, the energy of the true bound state has to be also reproduced.

However, in our work we analyze short-range correlations emerging at a domain wall of the phase-separated state where the strength of the beyond-contact attraction is treated as a free parameter.
Intuitively, length scale $\ell$ should reflect a range of the real interatomic potential, but in our analysis it is much larger.
Thus, the scaling with decreasing $\ell$ needs to be addressed.
Many-body simulations we performed are not attainable for significantly larger number of sites that would allow for studies of the system behavior when $\ell$ is smaller.
Nevertheless, we have found no evidence of qualitative differences when $\ell$ was increased twice and other parameters were adequately rescaled.
This may suggest that our findings hold also if smaller---and more realistic in unconstrained atomic systems---values of $\ell$ were to be used.

\textit{Conclusions and outlook}---We have proposed to utilize the three-delta interaction potential to study competition between many-body effects, such as pairing and ferromagnetism, in an ultracold two-component Fermi gas.
We have analyzed the dynamical stability of the experimentally inspired artificial domain structure and have found three stabilization mechanisms.
First one is due to usual mean-field repulsion of the components, while the other two involve many-body processes.
The process dominating in the presence of strong attraction can be associated with the stability of the super-Tonks-Girardeau gas, while in the regime of moderate attraction, the stabilization is due to the molecule formation at the interlayer between the components.
As a future line of work, the three-delta potential can become a powerful asset in the many-body studies of systems where repulsive and attractive interactions compete to form novel quantum phases.

\begin{acknowledgments}
\textit{Acknowledgments}---The authors would like to thank Titas Chanda, Krzysztof Jachymski and Vicky C. Arizona for fruitful discussions. 
A. S. acknowledges the support from Olle Engkvists stiftelse.
M. \L{}. acknowledges the support from the (Polish) National Science Center Grant 2018/31/N/ST2/01429. 
P. T. G. is financed from the (Polish) National Science Center Grants 2018/29/B/ST2/01308 and 2020/36/T/ST2/00065. 
K. Rz. is supported from the (Polish) National Science Center Grant 2018/29/B/ST2/01308.
 Center for Theoretical Physics of the Polish Academy of Sciences is a member of the National Laboratory of Atomic, Molecular and Optical Physics (KL FAMO).
\end{acknowledgments}
\bibliography{library}
\newpage


\section*{Supplementary material}

# Many-body molecule formation at a domain wall in a one-dimensional strongly interacting ultracold Fermi gas: Supplemental materials

Andrzej Syrwid,[1] Maciej Łebek,[2,3] Piotr T. Grochowski,[2,4,*] and Kazimierz Rzążewski[2]

[1]*Department of Physics, The Royal Institute of Technology, Stockholm SE-10691, Sweden*
[2]*Center for Theoretical Physics, Polish Academy of Sciences, Aleja Lotników 32/46, 02-668 Warsaw, Poland*
[3]*Institute of Theoretical Physics, University of Warsaw, Pasteura 5, 02-093 Warsaw, Poland*
[4]*ICFO - Institut de Ciències Fotòniques, The Barcelona Institute of Science and Technology, 08860 Castelldefels (Barcelona), Spain*

(Dated: May 2, 2021)

## I. BOUND STATES IN 1+1 PROBLEM

In this section we investigate a one-dimensional problem of two particles interacting via a three-delta potential and determine a criterion for a bound state formation in the absence of external confinement. The Hamiltonian of our system reads

$$H = -\frac{\hbar^2}{2m}\left(\partial_x^2 + \partial_y^2\right) + W(x-y) \quad (1)$$

with $W(x) = c_0 \delta(x) + c_\ell \delta(x-\ell) + c_\ell \delta(x+\ell)$ and we assume that $c_0 > 0$ and $c_\ell < 0$. We decouple center of mass (CM) and relative motion by introducing new coordinates $R = (x+y)/\sqrt{2}$ and $r = (x-y)/\sqrt{2}$ obtaining $H = H_{\rm CM} + H_{\rm rel}$, where $H_{\rm CM} = -\frac{\hbar^2}{2m}\partial_R^2$ and $H_{\rm rel} = -\frac{\hbar^2}{2m}\partial_r^2 + W'(r)$. Here $W'(r)$ denotes the three-delta potential with $\ell' = \ell/\sqrt{2}$ and $c'_{0,\ell} = c_{0,\ell}/\sqrt{2}$. We focus on $H_{\rm rel}$ and seek for localized eigenstates (bound states) describing a pair bound due to attractive part of $W'(r)$. In general the potential may have at most two bound states - one symmetric ($+$) and one antisymmetric ($-$). The problem is solved by imposing the following standard continuity and $\delta$-potential-related conditions

$$\psi_\pm(r) = \begin{cases} A_\pm e^{kr} & r < -\ell' \\ B_\pm e^{kr} + C_\pm e^{-kr} & -\ell' \leq r < 0 \\ \pm C_\pm e^{kr} \pm B_\pm e^{-kr} & 0 \leq r < \ell' \\ \pm A_\pm e^{-kr} & r \geq \ell' \end{cases}, \quad (2)$$

where $k = \sqrt{2m|E|}/\hbar$, with $E$ being the binding energy. The abovementioned conditions lead to the following relations $B_+ = C_+(2 - u_0/t)/(2 + u_0/t)$, $A_+ = -2C_+ t e^{2t}/u_\ell$ for symmetric solutions and $B_- = -C_-$, $A_- = -2C_- t e^{2t}/u_\ell$ for antisymmetric ones, where the dimensionless parameters $u_0 = mc_0\ell/\hbar^2$, $u_\ell = mc_\ell\ell/\hbar^2$ and $t = k\ell'$ were introduced. The energies $E$ can be determined thanks to relations

$$\begin{cases} 2\cosh t + \frac{u_0}{t}\sinh t + (2t + u_0)\frac{e^t}{u_\ell} = 0 & \text{for } (+) \\ te^t = -u_\ell \sinh t & \text{for } (-) \end{cases}. \quad (3)$$

* piotr@cft.edu.pl

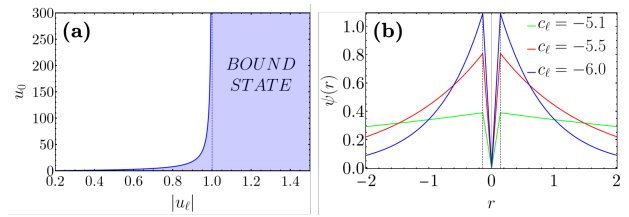

FIG. 1. Panel (a): Diagram of a bound state existence for the system described by (1). Shaded area represents the region of dimensionless parameters $u_0$ and $u_\ell$ where a bound state exists. Dashed line mark the critical value $u_\ell^{\rm crit} = -1$ below which the three-delta potential has a bound state no matter how strong the short range repulsion $u_0$ is. Panel (b): Bound state relative wavefunctions of the three-delta potential in the regime of strong contact repulsion. Note that with increasing $|c_\ell|$ the particles localize more and more eagerly at $r = \pm \ell'$ (marked with dashed lines). The relative wavefunctions vanish at $r = 0$ due to the strong point-like repulsion. Results were obtained for $\hbar = m = 1$, $c_0 = 5000$, and $\ell = 0.2$.

As the lowest energy state is always symmetric, our potential hosts a bound state only if (3) has a positive solution. Assuming a very strong contact repulsion $c_0 \to \infty$, we take the limit $t \to 0^+$ in (3) and find the critical value of attractive coupling $u_\ell^{\rm crit} = -1$ below which there exists a bound state in the system. While the regime of bound state existence is shown in Fig. 1(a), exemplary shapes of the corresponding bound state wavefunctions can be found in Fig. 1(b).

## II. 1+1 PROBLEM IN A HARMONIC TRAP

Let us now study the problem of two initially separated particles interacting via three-delta potential. We show that exact solutions provides us with a lot of insights about the time evolution and the structure of the energy spectrum. Additionally, it will serve as a benchmark for numerical methods employed when investigating larger systems.

## A. Solutions

Here, applying methods similar to [1, 2], we solve the problem that may be viewed as an extension of the problem analyzed in [1]. Namely, we consider the Hamiltonian

$$H = -\frac{\hbar^2}{2m}\left(\partial_x^2 + \partial_y^2\right) + \frac{1}{2}m\omega^2(x^2+y^2) + W(x-y). \quad (4)$$

As in the previous section we introduce $R$ and $r$ variables obtaining $H = H_{\text{CM}} + H_{\text{rel}}$, where $H_{\text{CM}}$ describes the standard Hamiltonian of harmonic oscillator and thus we focus only on $H_{\text{rel}} = -\frac{\hbar^2}{2m}\partial_r^2 + \frac{m\omega^2}{2}r^2 + W'(r)$. In the absence of potential $W'(r)$ the corresponding Schrödinger equation describes a standard harmonic oscillator in 1D, i.e., $E\psi(r) = -\frac{\hbar^2}{2m}\partial_r^2\psi(r) + \frac{m\omega^2}{2}r^2\psi(r)$. The most general solutions of the problem can be grouped into symmetric and antisymmetric ones and cast into the following forms

$$\psi_+(r) = \alpha_+ \tilde{F}_{\frac{\sigma}{2},\frac{1}{2}}(r) + \beta_+ \tilde{U}_{\frac{\sigma}{2},\frac{1}{2}}(r), \quad (5)$$

$$\psi_-(r) = \alpha_- \frac{r}{d}\tilde{F}_{\frac{\kappa}{2},\frac{3}{2}}(r) + \beta_- \frac{r}{d}\tilde{U}_{\frac{\kappa}{2},\frac{3}{2}}(r), \quad (6)$$

where $d = \sqrt{\hbar/m\omega}$ is the oscillator length, and the energy-related coefficients $\kappa = 3/2 - E/\hbar\omega$ and $\sigma = 1/2 - E/\hbar\omega$ [2, 3]. Additionally, for convenience we introduced functions $\tilde{F}_{\xi,\zeta}(r) = \exp(-r^2/2d^2)\,_1F_1(\xi;\zeta;r^2/d^2)$ and $\tilde{U}_{\xi,\zeta}(r) = \exp(-r^2/2d^2)U(\xi,\zeta,r^2/d^2)$, where $U$ and $_1F_1$ are confluent hypergeometric functions of the first and second type, respectively. It is important for further analysis to list the most important properties of $_1F_1$ and $U$ in order to construct physical eigenstates (taking delta functions into the account) similarly to the case without confinement [2, 3]. If $\sigma = -n$ with $n = 0, 1, 2, \ldots$, then $_1F_1\left(-\frac{n}{2};\frac{1}{2};x^2\right) \sim H_{2n}(x)$ and for $\kappa = -n$ ($n = 0, 1, 2, \ldots$) we get $_1F_1\left(-\frac{n}{2};\frac{3}{2};x^2\right) \sim H_{2n+1}(x)$. In this way we retrieve free oscillator solutions out of general forms (5) and (6). However, when interactions shift values of energies so that $\sigma$ and $\kappa$ are no longer nonpositive integer numbers, the functions $\tilde{F}_{\frac{\sigma}{2},\frac{1}{2}}(r)$ and $\frac{r}{d}\tilde{F}_{\frac{\kappa}{2},\frac{3}{2}}(r)$ become nonnormalizable. This means that in the region where $|r| > \ell'$ we have solutions with $\alpha^{(+)} = 0$ ($\alpha^{(-)} = 0$) in the symmetric (antisymmetric) case. On the other hand, in that case functions $\tilde{U}_{\frac{\sigma}{2},\frac{1}{2}}(r)$ and $\frac{r}{d}\tilde{U}_{\frac{\kappa}{2},\frac{3}{2}}(r)$ decay sufficiently fast assuring normalizability of the wave function. Now, regarding behavior of the wave functions at the origin, functions $\tilde{F}_{\frac{\sigma}{2},\frac{1}{2}}(r)$ and $\frac{r}{d}\tilde{F}_{\frac{\kappa}{2},\frac{3}{2}}(r)$ are continuous and have continuous derivatives at $r = 0$. The situation is slightly different for $\tilde{U}_{\frac{\sigma}{2},\frac{1}{2}}(r)$ and $\frac{r}{d}\tilde{U}_{\frac{\kappa}{2},\frac{3}{2}}(r)$ which are continuous at $r = 0$ but reveal discontinuities in derivatives at this point, which we use to fulfill the abovementioned Dirac delta conditions. These properties motivate us to propose solutions in the form

$$\psi_+(r) = \begin{cases} A_+\tilde{U}_{\frac{\sigma}{2},\frac{1}{2}}(r) & |r| \geq \ell' \\ B_+\tilde{F}_{\frac{\sigma}{2},\frac{1}{2}}(r) + C_+\tilde{U}_{\frac{\sigma}{2},\frac{1}{2}}(r) & |r| < \ell' \end{cases}, \quad (7)$$

$$\psi_-(r) = \begin{cases} A_-\frac{r}{d}\tilde{U}_{\frac{\kappa}{2},\frac{3}{2}}(r) & |r| \geq \ell' \\ B_-\frac{r}{d}\tilde{F}_{\frac{\kappa}{2},\frac{3}{2}}(r) & |r| < \ell' \end{cases}. \quad (8)$$

Now, we proceed similarly as in the case with no confinement, i.e., we impose continuity and $\delta$-potential-related conditions at $r = 0, \pm\ell'$. Let us start with the symmetric solutions, where continuity of the wave function at $r = \pm\ell'$ implies

$$A_+ U\left(\frac{\sigma}{2},\frac{1}{2},\frac{\ell'^2}{d^2}\right) = B_+ {}_1F_1\left(\frac{\sigma}{2};\frac{1}{2};\frac{\ell'^2}{d^2}\right) + C_+ U\left(\frac{\sigma}{2},\frac{1}{2},\frac{\ell'^2}{d^2}\right) \quad (9)$$

and $\delta$-induced condition at $r = 0$ yields (see [3] for details)

$$C_+\hbar\omega d\frac{\sqrt{\pi}\sigma}{\Gamma\left(1+\frac{\sigma}{2}\right)} + c'_0\left[B_+ + C_+\frac{\sqrt{\pi}}{\Gamma\left(\frac{1}{2}+\frac{\sigma}{2}\right)}\right] = 0. \quad (10)$$

Applying the same conditions but at $r = \pm\ell$ together with relations (9) and (10) one finds

$$B_+ = -\Xi_\sigma C_+, \quad (11)$$

$$A_+ = \frac{2\Xi_\sigma {}_1F_1\left(1+\frac{\sigma}{2};\frac{3}{2};\frac{\ell'^2}{d^2}\right) + U\left(1+\frac{\sigma}{2},\frac{3}{2},\frac{\ell'^2}{d^2}\right)}{U\left(1+\frac{\sigma}{2},\frac{3}{2},\frac{\ell'^2}{d^2}\right) - 2\frac{c'_\ell}{\hbar\omega\ell'\sigma}U\left(\frac{\sigma}{2},\frac{1}{2},\frac{\ell'^2}{d^2}\right)}C_+, \quad (12)$$

where $\Xi_\sigma = \sqrt{\pi}\left[\hbar\omega d\sigma/c'_0\Gamma\left(1+\frac{\sigma}{2}\right) + 1/\Gamma\left(\frac{1}{2}+\frac{\sigma}{2}\right)\right]$ was introduced for convenience. By plugging these equalities into the continuity condition, we determine the equation for $\sigma$ (directly related to energies)

$$\frac{\sigma U\left(\frac{\sigma}{2},\frac{1}{2},\frac{\ell'^2}{d^2}\right)}{U\left(\frac{\sigma}{2},\frac{1}{2},\frac{\ell'^2}{d^2}\right) - \Xi_\sigma {}_1F_1\left(\frac{\sigma}{2};\frac{1}{2};\frac{\ell'^2}{d^2}\right)} \quad (13)$$

$$= \frac{\sigma U\left(1+\frac{\sigma}{2},\frac{3}{2},\frac{\ell'^2}{d^2}\right) + \frac{2c'_\ell}{\hbar\omega\ell'}U\left(\frac{\sigma}{2},\frac{1}{2},\frac{\ell'^2}{d^2}\right)}{2\Xi_\sigma {}_1F_1\left(1+\frac{\sigma}{2};\frac{3}{2};\frac{\ell'^2}{d^2}\right) + U\left(1+\frac{\sigma}{2},\frac{3}{2},\frac{\ell'^2}{d^2}\right)}.$$

When dealing with antisymmetric solutions, the corresponding wave functions are not affected by $\delta$-potential at $r = 0$ and thus it is enough to focus on $r = \pm\ell'$ cases, where from the continuity and $\delta$-related conditions we find

$$A_- U\left(\frac{\kappa}{2},\frac{3}{2},\frac{\ell'^2}{d^2}\right) = B_- {}_1F_1\left(\frac{\kappa}{2};\frac{3}{2};\frac{\ell'}{d^2}\right), \quad (14)$$

$$A_- = -\frac{\kappa\left[\frac{A_-}{2}U\left(1+\frac{\kappa}{2},\frac{5}{2},\frac{\ell'^2}{d^2}\right) + \frac{B_-}{3}{}_1F_1\left(1+\frac{\kappa}{2};\frac{5}{2};\frac{\ell'^2}{d^2}\right)\right]}{c'_\ell U\left(\frac{\kappa}{2},\frac{3}{2},\frac{\ell'^2}{d^2}\right)}. \quad (15)$$

Finally, $\kappa$ gives the energies of antisymmetric states and can be determined from

$$-\kappa = \quad (16)$$

$$\frac{\frac{2c'_\ell}{\hbar\omega\ell'}U\left(\frac{\kappa}{2},\frac{3}{2},\frac{\ell'^2}{d^2}\right){}_1F_1\left(\frac{\kappa}{2};\frac{3}{2};\frac{\ell'^2}{d^2}\right)}{{}_1F_1\left(\frac{\kappa}{2};\frac{3}{2};\frac{\ell'^2}{d^2}\right)U\left(1+\frac{\kappa}{2},\frac{5}{2},\frac{\ell'^2}{d^2}\right) + \frac{2}{3}U\left(\frac{\kappa}{2},\frac{3}{2},\frac{\ell'^2}{d^2}\right){}_1F_1\left(1+\frac{\kappa}{2};\frac{5}{2};\frac{\ell'^2}{d^2}\right)}.$$

Our eigenproblem is now solved. Energies are readily obtained as numerical solutions of (13) and (16). Wave

functions are determined from Eqs. (11), (12) and (14) together with the normalization condition. Ground states for different parameters are presented in Fig.2. In summary, let us note that full eigenstates of the two body problem can be cast into the following form

$$\Psi_{n,\mu}(x,y) = \phi_n\left(\frac{x+y}{\sqrt{2}}\right)\psi_\mu\left(\frac{x-y}{\sqrt{2}}\right), \qquad (17)$$

where $\phi_n$, $n = 0, 1, 2, \ldots$ is a harmonic oscillator eigenstate whereas $\psi_\mu$ represents an eigenstate of $H_{\rm rel}$. Here $\mu$ denotes either $\kappa$ or $\sigma$ depending on the symmetry of a given state in the relative coordinate $r$.

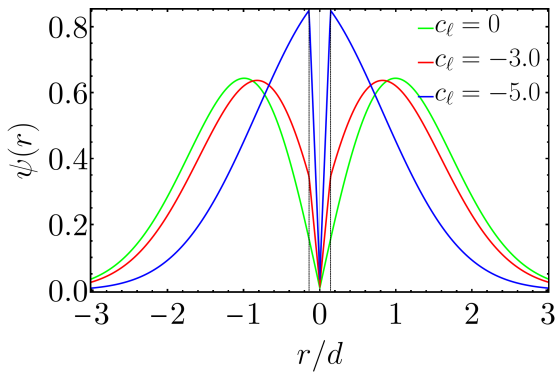

FIG. 2. Ground states of the relative Hamiltonian $H_{\rm rel}$ for $\ell/d = 0.2$, $c_0/\hbar\omega d = 200$ and different $c_\ell$. Similarly to the case without confinement, the solution becomes more and more localized at $r = \pm \ell'$ as we increase $|c_\ell|$.

### B. Super Tonks-Girardeau-like behavior

As discussed in the main text, the system dynamics in the presence of strong finite-range attraction turns out to be very similar to the one observed in the case of a purely repulsive contact interaction. We argue that this counterintuitive behavior of the system can be related to the phenomenon of the super Tonks-Girardeau gas (sTG) [4–7].

The sTG is a highly excited, metastable state of matter where the particles do not form bound states despite a strong interparticle attraction. Such a state has a large overlap with the ground state of infinitely repulsive Tonks-Girardeau gas (TG). This phenomenon was understood over ten years ago in the Lieb-Liniger model [5]. It turns out that a sudden quench of interactions from strongly repulsive to strongly attractive does not drive the system out of the equilibrium. A trace of such a behavior can be observed already for two confined particles interacting via contact potential [1, 6]. As found in Ref. [1], the spectrum of the relative Hamiltonian in the strongly attractive regime equals the spectrum of the strongly repulsive one with an additional very deep bound state.

In this section we show that the situation is similar in the case of the three-delta interaction. Let us consider the limit of $c_0/\hbar\omega d \to \infty$ in Eqs. (13) and (16), where the spectrum becomes double-degenerate and all the states can be grouped into pairs consisting of one symmetric and antisymmetric characterized by equal energies. The resulting eigenenergies of the relative motion as a function of $c_\ell$ for fixed $\ell$ are presented in Fig. 3. We find that when $c_\ell/\hbar\omega d \to -\infty$ the spectrum equals the spectrum of the system in the limit $c_\ell/\hbar\omega d \to \infty$ with additional two infinitely deep bound states (symmetric and antisymmetric).

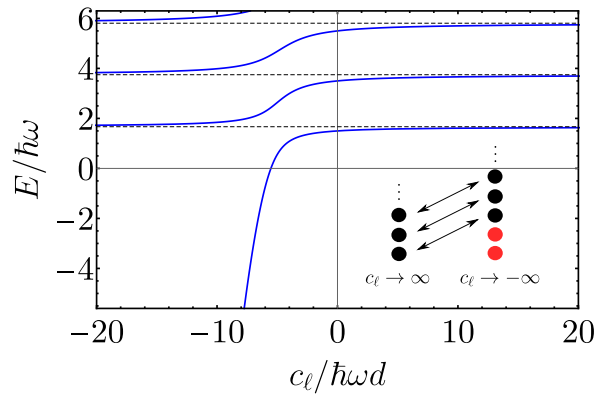

FIG. 3. The spectrum of $H_{\rm rel}$ in the limit $c_0/\hbar\omega d \to \infty$ for $\ell/d = 0.2$ as a function of $c_\ell$. Due to the degeneracy, each curve corresponds to two states: symmetric and antisymmetric. There are two bound states (symmetric and antisymmetric) for sufficiently negative $c_\ell/\hbar\omega d$. The inset presents a pictorial representation of the relationship between spectra for $c_\ell/\hbar\omega d \to \infty$ and $c_\ell/\hbar\omega d \to -\infty$, where red dots represent bound states.

In general, the dynamics of the system is given by

$$\Psi(x,y;t) = \sum_{n,\mu} e^{-\frac{i}{\hbar}(E_n+E_\mu)t} C_{n,\mu} \Psi_{n,\mu}(x,y) \qquad (18)$$

where $E_n$ denotes the $n$-th harmonic oscillator eigenenergy, $E_\mu$ is the $\mu$-th eigenenergy of $H_{\rm rel}$ and $C_{n,\mu} = \langle\Psi_{n,\mu}|\Psi_0\rangle$, where $\Psi_0$ and $\Psi_{n,\mu}$ represent the initial state and the eigenstate of the form (17), respectively. Let us focus on the case $c_\ell/\hbar\omega d \to -\infty$. As the initial state, $\Psi_0$, corresponds to two initially separated atoms, the wave function is zero when $x = y$ and is still very small for $|x-y| = \ell$. On the other hand, the bound states, $\Psi_b$, are tightly localized at $|x-y| = \ell$ and almost zero elsewhere. Consequently, $\langle\Psi_0|\Psi_b\rangle \to 0$ and both symmetric and antisymmetric bound states effectively do not contribute to the dynamics (18). The remaining states that do have a meaningful overlaps, are at the same time eigenstates of the system with $c_\ell/\hbar\omega d \to \infty$. As a result, the dynamics of the system with strong finite-range attraction closely follows the dynamics of the system with strongly repulsive potential $c_\ell/\hbar\omega d \to \infty$. We will elaborate on the mechanisms presented here in subsequent work [8].

## III. LOCV APPROXIMATION

In this section we go beyond the two-body problem and consider a uniform, one-dimensional Fermi gas of $N_\uparrow$ spin-$\uparrow$ and $N_\downarrow$ spin-$\downarrow$ atoms characterized by mass $m$. The balanced gas, $N_\uparrow = N_\downarrow = M = N/2$, is described by the following Hamiltonian

$$H = -\frac{\hbar^2}{2m}\sum_{i,\sigma}\partial^2_{x_i^\sigma} + \sum_{i,j} W(x_i^\uparrow - x_j^\downarrow), \quad (19)$$

where $x_i^\sigma$ denote spatial coordinates of spin-$\sigma$ atoms and $W(r)$ is a short-range two-body interaction potential between particles belonging to different spin components, with no spin-flipping allowed. Here, the total wave function is approximated by means of the Jastrow-Slater Ansatz [9]:

$$\Psi(\{x_i^\uparrow\},\{x_j^\downarrow\}) = J(\{x_i^\uparrow\},\{x_j^\downarrow\})\prod_\sigma \mathcal{D}_\sigma(\{x_i^\sigma\}), \quad (20)$$

where $\mathcal{D}_\uparrow$ and $\mathcal{D}_\downarrow$ are Slater determinants:

$$\mathcal{D}_\sigma(\{x_i^\sigma\}) = \frac{1}{\sqrt{M!}}\begin{vmatrix} \varphi_1^\sigma(x_1^\sigma) & \cdots & \varphi_1^\sigma(x_M^\sigma) \\ \cdot & & \cdot \\ \cdot & & \cdot \\ \cdot & & \cdot \\ \varphi_M^\sigma(x_1^\sigma) & \cdots & \varphi_M^\sigma(x_M^\sigma) \end{vmatrix}, \quad (21)$$

of plane waves $\varphi_i^\sigma$, and $J$ is a Jastrow factor

$$J(\{x_i^\uparrow\},\{x_j^\downarrow\}) = \prod_{i,j} f(|x_i^\uparrow - x_j^\downarrow|), \quad (22)$$

with $f$ being a Jastrow function that accounts for a two-body relative wave function of opposite spin atoms. Such an Ansatz is not valid for general Fermi systems, e.g. neutron matter, however it proved to be a good approximation for cold atomic systems in which short-range interactions are dominant [10].

We assume that the Jastrow factor slightly modifies the whole wave function at short relative distances only and that the parameter $F_{ij} = f^*(|x_i^\uparrow - x_j^\downarrow|)f(|x_i^\uparrow - x_j^\downarrow|) - 1$ is small. Thus, we can expand $|J|^2 = 1 + \sum_{i,j} F_{ij} + O(F^2)$.

Let us proceed to evaluate the normalization of the wave function. By the orthonormality of plane waves, we find that up to the first order in $F_{ij}$

$$\int \mathrm{d}x \int \mathrm{d}y\, [f^*(x-y)f(x-y) - 1]\, n_\uparrow n_\downarrow = 1, \quad (23)$$

which under the transformation $r = x - y$, $R = x + y$ can be cast into $n\int \mathrm{d}r |f(r)|^2 = N$, where $n_\sigma = n/2$ denotes the $\sigma$-particle density.

Now, we need to introduce physical assumptions on the behavior of the Jastrow function. We demand that for long separations between the atoms, $J$ asymptotically tends to 1 not to affect the long range behavior of the wave function. Even more than that, we introduce the healing length $d$ that is of order of average interparticle separation $1/n$, beyond which the Jastrow function does not modify $\Psi$. It expresses the intuition that on average only the nearest neighbors of opposite spins are correlated and the correlations with atoms at longer distances are negligible. This constraint and the requirement for smoothness at the healing distance,

$$f(|r| \geq d) = 1, \qquad f'(d) = f'(-d) = 0, \quad (24)$$

are two out of three constraining conditions. The last one comes from the normalization of the Jastrow factor. We assume that $f$ deviates from 1 only inside the sphere with radius $d$, $K(d)$, and on average there is only a single $\uparrow$-$\downarrow$ pair inside $K(d)$. Therefore, starting with $n\int \mathrm{d}r|f(r)|^2 = N$, assuming $L$ to be the system size one rewrites

$$\frac{n}{2}\left[\int_{K(d)} \mathrm{d}r\, |f(r)|^2 + \int_{V-K(d)} \mathrm{d}r\, 1\right] = \frac{N}{2}, \quad (25)$$

which results in

$$\frac{n}{2}\left[\int_{K(d)} \mathrm{d}r\, |f(r)|^2 + \frac{N/2 - 1}{N/2}V\right] = \frac{N}{2}. \quad (26)$$

It gives the last constraint

$$\frac{n}{2}\int_{K(d)} \mathrm{d}r\, |f(r)|^2 = 1. \quad (27)$$

These three constraints are needed in variational calculations in order to reproduce the experimental data [10, 11]. Unconstrained calculations were utilized to study nuclear matter, however each time Jastrow function tended to be unphysically long-ranged. On the other hand, constrained studies provided extremely good fit to the experimental data [12, 13]. Note that introduction of these constraints will also imply some kind of the effective two-body interaction. If the Jastrow function were to satisfy the usual two-body Schrödinger equation for the relative wave function, it would not tend to 1 for long distances. Therefore, some kind of alternation of interaction is needed in order to satisfy the constraints, which will be addressed later.

We will now proceed to evaluate the energy within the Slater-Jastrow Ansatz. The noninteracting contribution comes from the kinetic energy of the orbitals and in large atom number limit is equal to $E_0/V = nE_F/3$, where the Fermi energy $E_F = \hbar^2\pi^2 n^2 V/8m$. The interaction part comes as a two-body cluster expansion of the total energy that takes into account both the kinetic energy from the Jastrow function and the influence of a bare two-body potential. Namely,

$$E_\mathrm{int} = \left\langle \hat{H} \right\rangle - E_0 \approx \int \mathrm{d}x\mathrm{d}y\, f^*(|x-y|)$$
$$\times \left[-\frac{\hbar^2}{2\mu}\partial^2_{x-y} + W(x-y)\right] f(|x-y|)\, n_\uparrow n_\downarrow, \quad (28)$$

$$\frac{E_\mathrm{int}}{V} = n_\uparrow n_\downarrow \int \mathrm{d}r\, f^*(r)\left(-\frac{\hbar^2}{2\mu}\partial^2_r + W(r)\right)f(r). \quad (29)$$

Now, Eq. (29) needs to be extremized with respect to variations of $f$. However, in order for $f$ to satisfy the constraints, two-body interaction needs to be altered. The easiest way to do so is by adding additional constant external potential $\lambda$ in such a way that $W \to W - \lambda$ [12]. It is not the only way of renormalization, however it was shown to provide reliable results and to be equivalent to so called Moszkowski-Scott separation [14] and Brueckner theory [15]. This procedure can be understood as an effect of average pressure of further neighbors and sometimes it is interpreted as a contribution of two-body correlations to average field felt by a given atom.

Therefore, to find a minimum, one needs to solve

$$0 = \delta \int dr\, f^*(r) \left( -\frac{\hbar^2}{2\mu}\partial_r^2 + W(r) - \lambda \right) f(r). \quad (30)$$

It yields two-body Schrödinger-like equation

$$-\left[ \frac{\hbar^2}{2\mu}\partial_r^2 + W(r) \right] f(r) = \lambda f(r), \quad (31)$$

that needs to be solved with constraints

$$f(d) = f(-d) = 1, \quad (32)$$
$$f'(d) = f'(-d) = 0, \quad (33)$$
$$\frac{n}{2}\int_{-d}^{d} |f|^2(r)\, dr = 1. \quad (34)$$

The interaction energy is then

$$\frac{E_{\text{int}}}{V} = n_\uparrow n_\downarrow \int dr\, f^*(r)\left[ -\frac{\hbar^2}{2\mu}\partial_r^2 + W(r)\right] f(r) = \frac{n\lambda}{2}. \quad (35)$$

The final result, $E_{\text{int}}/V = n\lambda/2$, indicates that on average we have $n/2$ pairs, each one contributing energy $\lambda$.

The LOCV approximation proved to provide reliable experimental predictions in many settings, from dense nuclear matter to ultracold quantum mixtures. However, it was mostly utilized in three-dimensional geometries, in which the level of its accuracy matches state-of-the-art quantum Monte Carlo calculations [10, 13]. Analysis of low dimensional systems was scarce, but the preliminary results in two dimensional Fermi gas suggest that LOCV fares worse in such geometries. The probable reason is a larger role of quantum fluctuations in lower dimensions and resulting need to go beyond the lowest order in the cluster expansion to match the more sensitive approaches. Nevertheless, the qualitative match can be expected as LOCV should reproduce leading order behavior of systems considered.

### A. Three-delta potential in LOCV approximation

We now consider a specific potential consisting of three delta contributions

$$W(x) = c_0\delta(x) + c_\ell\delta(x-\ell) + c_\ell\delta(x+\ell), \quad (36)$$

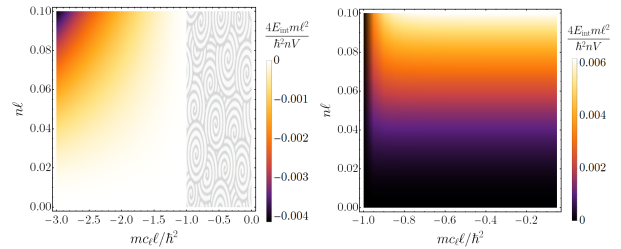

FIG. 4. Interaction energy at the attractive (left panel) and the repulsive (right panel) branch of the many-body spectrum with the three-delta potential within the LOCV approximation as a function of $n\ell$ and $mc_\ell\ell/\hbar^2$. The bound state is supported only when $mc_\ell\ell/\hbar^2 \lesssim -1$.

where we choose the contact interaction strength $c_0$ to be large and positive to mimic the repulsive core of realistic interparticle potential. On the other hand, beyond-contact part, $c_\ell$, is chosen to be small and negative to support a weakly bound state and provide the simplest extension of a contact part towards full van der Waals forces.

In a three-dimensional two-component Fermi mixture, the contact repulsion given by the Fermi-Huang pseudopotential $\propto \delta(r)\frac{\partial}{\partial r}r$ reproduces both the weakly bound pairs (attractive branch of the many-body system) and the repulsive, anticorrelated, phase-separated Fermi clouds (repulsive branch). This is not the case in a one-dimensional space as repulsive contact interactions do not support a bound state.

To simplify our considerations, we will work in the regime of infinite repulsion, $c_0 \to \infty$, yielding $f(0) = 0$.

We compute the interaction energy within the LOCV approximation solving Eqs. (31)–(34). If the bound state is considered, $\lambda = -\hbar^2\kappa^2/2m$ is negative. Then, we assume the Jastrow function to take form

$$f(x) = \begin{cases} A\sinh(\kappa x) & : x \in [0, \ell] \\ B\sinh(\kappa x) + D\cosh(\kappa x) & : x \in (\ell, d] \\ 1 & : x > d, \end{cases} \quad (37)$$

and needs to be self-consistently solved for $A, B, D, d, \kappa$ with constraints (33)–(34). It simplifies to the following two equations

$$\frac{2}{na} = \cosh^2(\kappa_1 - \kappa_2)\left[\frac{\coth(\kappa_1)}{\kappa_1} - \operatorname{csch}^2(\kappa_1)\right]$$
$$+ \frac{1}{2\kappa_1}\left[2\kappa_2 - 2\kappa_1 + \sinh(2\kappa_2 - 2\kappa_1)\right], \quad (38)$$

$$\tanh(\kappa_2) = \frac{1}{\tanh(\kappa_1)} + \frac{\kappa_1}{\lambda_2}\frac{1}{\sinh^2(2\kappa_1)}, \quad (39)$$

where we define $\kappa_1 = \kappa\ell$, $\kappa_2 = \kappa d$, $\lambda_2 = mc_\ell\ell/\hbar^2$. The interaction energy density then equals

$$\frac{E_{\text{int}}}{V} = \frac{n}{2}\lambda = -\frac{n}{2}\frac{\hbar^2\kappa_1^2}{2m\ell^2}. \quad (40)$$

5

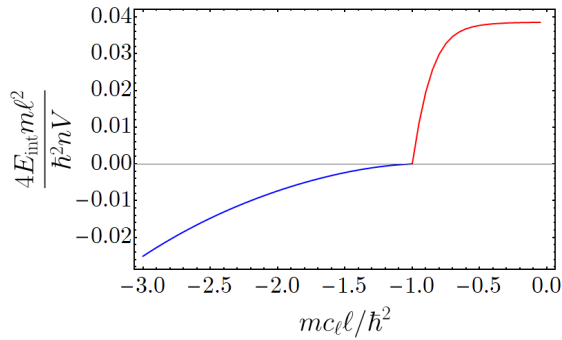

FIG. 5. Interaction energy of the ground state within the LOCV approximation for $n\ell = 0.25$. The sharp transition from positive to negative energy is clearly visible at $mc_\ell\ell/\hbar^2 \approx -1$ and indicates an appearance of the bound state in the system.

It is evaluated numerically and presented in Fig. 4(left).

Note that the bound state is not supported for every value of $c_\ell$. Most notably, numerical evaluation allows to write down the simple expression for the existence of the bound state

$$c_\ell \lesssim -\frac{\hbar^2}{m\ell}, \quad (41)$$

which does not depend on the gas density.

On the other hand, similar procedure for the first excited state (the repulsive branch) yields

$$\frac{2}{na} = [k_1 - \sin(k_1)\cos(k_1)][\cos(k_2)\cot(k_1) + \sin(k_2)]^2$$
$$+ \frac{1}{2k_1}[2k_2 - 2k_1 + \sin(2k_2 - 2k_1)], \quad (42)$$

$$\tan(k_2) = -\left[\frac{1}{\tan(k_1)} + \frac{k_1}{\lambda_2}\frac{1}{\sin^2(2k_1)}\right], \quad (43)$$

where $\lambda = \frac{\hbar^2 k^2}{2m}$ is positive and we define $k_1 = k\ell$, $k_2 = kd$, $\lambda_2 = mc_\ell\ell/\hbar^2$. The result is presented in Fig. 4(right).

The lowest energy branch of positive solutions to the constrained equations tends to 0 at $mc_\ell\ell/\hbar^2 \approx -1$, i.e. the interaction strength at which the bound state starts to be supported. In Fig. 5 the energy dependence on parameter $c_\ell$ for the many-body ground state at a given density is presented. The sharp transition is clearly visible, suggesting nontrivial behavior of the system at the critical value of $c_\ell$.

## IV. SHORT-RANGE ATOMIC POTENTIALS

### A. Scattering in quasi-one-dimensional geometry

Here we sum up the analysis of Ref. [16] where the authors show how finite-range corrections to the standard pointlike delta interaction can be modelled through the effective pseudopotential

$$W_p(x) = g_{1D}\left(1 + g'p^2\right)\delta(x), \quad (44)$$

where $g_{1D}$ denotes a one-dimensional mean field interaction strength, $g'$ is a parameter depending on the details of the physical situation, and following Ref. [16], $\hbar p$ represents a one-dimensional momentum operator.

The system under consideration involves ultracold atoms interacting via the attractive part of the van der Waals potential $W_{\text{vdW}}(r) = -C_6/r^6$ with a characteristic length (mean scattering length) $\bar{a} = \frac{2\pi}{\Gamma(1/4)^2}(2\mu C_6/\hbar^2)^{1/4}$, where $\mu$ is the reduced mass of the atomic pair [17] and $\Gamma$ is the Euler gamma function. It can be rephrased in terms of the van der Waals radius $R_{\text{vdW}}$ as $\bar{a} = 0.955978 R_{\text{vdW}}$.

In the presence of tight harmonic confinement $\frac{1}{2}\mu\omega^2\rho^2$ $\rho = \sqrt{y^2 + z^2}$ and $\omega$ denotes trapping frequency, the usual Fermi-Huang pseudopotential can be generalized to account for an energy dependence

$$W_{\text{FH}}(\mathbf{r}) = -\frac{2\pi\hbar^2}{\mu}\frac{\tan\delta_{3D}(k)}{k}\delta(\mathbf{r})\frac{\partial}{\partial r}r, \quad (45)$$

where $\hbar^2 k^2/2\mu$ is the kinetic energy of the relative motion and $\delta_{3D}(k)$ is the the phase shift due to interactions. Consequently, the corresponding energy-dependent scattering length reads $a_{3D}(k) = -\tan\delta_{3D}(k)/k$.

One notes that the full three-dimensional scattering analysis can be reduced to the 1D problem in the s-wave channel with the pseudopotential [16, 18–21]

$$W_{1D} = g_{1D}(p)\delta(x), \quad (46)$$

where

$$g_{1D}(p) = -\frac{\hbar^2}{\mu}p\tan[\delta_{1D}(p)]. \quad (47)$$

Here, $\delta_{1D}(p)$ is a 1D phase shift depending on 1D momentum $\hbar p$, where $k^2 = p^2 + 2/d_\perp^2$ with $d_\perp = \sqrt{\hbar/\mu\omega}$.

Expanding (47) in small $p$ up to the second order one finds

$$g_{1D}(p) \approx g_{1D}\left(1 + g'p^2\right), \quad (48)$$

with

$$\frac{\mu}{\hbar^2}g_{1D} = \frac{2}{d_\perp}\left(\frac{d_\perp}{a_{3D}} - C - \frac{r_{3D}}{d_\perp}\right)^{-1}, \quad (49)$$

$$g' = \frac{d_\perp}{2}\frac{r_{3D} - \tilde{C}d_\perp}{\frac{d_\perp}{a_{3D}} - C - \frac{r_{3D}}{d_\perp}}, \quad (50)$$

where $\tilde{C} = \zeta(3/2)/8 \approx 0.3265$ and $C = -\zeta(1/2) \approx 1.46035$. The effective range $r_{3D}$ can be determined from the expansion

$$k\cot[\delta_{3D}(k)] = -\frac{1}{a_{3D}} + \frac{1}{2}r_{3D}k^2 + O(k^2). \quad (51)$$

In vicinity of the Feshbach resonance, the effective range of a single-channel van der-Waals potential can be determined analytically [22–24],

$$r_{3D} = \frac{\Gamma(1/4)^4 \bar{a}}{6\pi^2}\left(1 - \frac{2\bar{a}}{a_{3D}} + \frac{2\bar{a}^2}{a_{3D}^2}\right) - 2R_*\left(1 - \frac{a_{bg}}{a_{3D}}\right)^2, \quad (52)$$

where $R_* = \hbar^2/(2\mu a_{bg}\Delta\delta\mu)$, $a_{bg}$ is the background scattering length away from the resonance, $\Delta$ denotes the resonance width and $\delta\mu$ is the magnetic moment difference between channels. This expression approximately holds for open channels, but can fail in the presence of the closed ones [24]. Then, it is more accurate to use formula

$$r_{3D}(B) \approx \frac{v + r_0(a_{3D}(B) - a_{ex})^2}{a_{3D}(B)^2}, \quad (53)$$

where the magnetic field dependence of $a_{3D}$ reads

$$a_{3D}(B) = a_{bg}\left(1 - \frac{\Delta}{B - B_0}\right), \quad (54)$$

with $B_0$ being the resonance point, and the fitting parameters $v$, $r_0$ and $a_{ex}$ depend on a particular resonance and can be evaluated through the numerical solutions of a full multi-channel Schrödinger equation.

### B. Mapping onto three-delta potential

We will now proceed to discretize effective interaction potential (44). The starting point is to consider two-body problem with a total wave function

$$\Psi_f(x, y) = \Psi_R(R)\Psi(r), \quad (55)$$

where $R = x + y$ describes the decoupled center-of-mass motion of two atoms, $r = x - y$ describes the relative motion, and $x$, $y$ represent positions of atoms. For the delta interaction $W_a(x-y) = g\delta(x-y-a)$, the interaction energy reads

$$E = \int \mathrm{d}x\mathrm{d}y\, \Psi_f^*(x,y) W_a(x-y)\Psi_f(x,y) = \frac{g}{2}|\Psi(a)|^2. \quad (56)$$

The position representation of potential (44) can be written as

$$W_p(x) = g_{1D}\left[\delta(x) + \frac{g'}{2}\left(\overleftarrow{\partial_x^2}\delta(x) + \delta(x)\overrightarrow{\partial_x^2}\right)\right]. \quad (57)$$

After discretizing the second derivative, i.e., $f''(x) \approx [f(x+\Delta x) + f(x-\Delta x) - 2f(x)](\Delta x)^{-2}$, the energy associated with the last term of the sum in (57) read

$$E = -\frac{g_{1D}g'}{4(\Delta x)^2} \qquad (58)$$
$$\times \int \mathrm{d}x[\Psi^*(x+\Delta x) + \Psi^*(x-\Delta x) - 2\Psi^*(x)]\delta(x)\Psi(x).$$

After realizing that

$$\frac{1}{2}\left[\Psi^*(x+\Delta x)\Psi(x) + \Psi(x+\Delta x)\Psi^*(x)\right]$$
$$= \Psi^*(x+\Delta x)\Psi(x+\Delta x) + O\left((\Delta x)^2\right), \quad (59)$$

one finds the interaction energy related to the finite-range correction

$$E = -\frac{g_{1D}g'}{2(\Delta x)^2}\left[\left|\Psi\left(\frac{\Delta x}{2}\right)\right|^2 + \left|\Psi\left(-\frac{\Delta x}{2}\right)\right|^2 - 2|\Psi(0)|^2\right]. \quad (60)$$

By Eq. (56), it corresponds to the following potential

$$W_b(x) = -\frac{g_{1D}g'}{(\Delta x)^2}\left[\delta\left(x + \frac{\Delta x}{2}\right) + \delta\left(x - \frac{\Delta x}{2}\right) - 2\delta(x)\right], \quad (61)$$

where $\Delta x$ can be arbitrarily chosen. It is convenient to denote $\Delta = \sqrt{|g'|}/\alpha$ with some arbitrary real number $\alpha > 0$. In result we reproduce the three-delta potential

$$W(x) = c_0\delta(x) + c_\ell\delta(x - \ell) + c_\ell\delta(x + \ell), \quad (62)$$

with

$$\begin{cases} c_0 = (1 + 2\alpha^2\,\mathrm{sgn}\,g')\,g_{1D} \\ c_\ell = -\alpha^2\,\mathrm{sgn}\,g'\,g_{1D} \\ \ell = \sqrt{|g'|}/2\alpha \end{cases}, \quad (63)$$

and inversely $g_{1D} = c_0 + 2c_\ell$, $g' = -4\ell^2 c_\ell/(c_0 + 2c_\ell)$.

Note that $g'$ and $g_{1D}$ can be uniquely determined by $c_0$, $c_\ell$ and $\ell$, while $\alpha$ provides an additional information how to perform inverse transformation. It is due to the fact that $g'$ and $g_{1D}$ describes only the scattering properties of the physical system. Effectively, $\alpha$ fine tunes the length scale introduced through the equality $\ell = \sqrt{|g'|}/2\alpha$ and can be evaluated by considering other properties of the two-body problem, such as the bound state energy.

## V. DISCRETIZATION OF THE CONTINUOUS MODEL

The Hamiltonian of the considered system reads

$$\hat{H} = \sum_{\sigma=\uparrow,\downarrow}\int\mathrm{d}x\,\hat{\psi}_\sigma^\dagger(x)\left[-\frac{\hbar^2}{2m}\partial_x^2 + \frac{1}{2}m\omega^2 x^2\right]\hat{\psi}_\sigma(x) \quad (64)$$
$$+ \int\mathrm{d}x\mathrm{d}y\,\hat{\psi}_\uparrow^\dagger(x)\hat{\psi}_\downarrow^\dagger(y)W(x-y)\hat{\psi}_\uparrow(x)\hat{\psi}_\downarrow(y),$$

where $\hat{\psi}_\sigma(x)$ denotes the canonical Fermi field annihilation operator satisfying standard fermionic anticommutation relations. We discretize the space into sites with spacing $\Delta$ and introduce new fermionic operators acting directly on lattice sites, i.e., $\hat{\psi}_\sigma(x_i) \to \sqrt{\Delta}\hat{a}_{\sigma,i}$ where $x_i$ represents a position of $i$-th lattice site.

In result we arrive at the Fermi-Hubbard model with beyond on-site interactions

$$\hat{H}_{FH} = \mathcal{T}\sum_{\sigma,i}\left(\hat{a}^\dagger_{\sigma,i}\hat{a}_{\sigma,i+1} - \hat{n}_{\sigma,i} + \text{h.c.}\right) + \mathcal{U}\sum_{\sigma,i} x_i^2 \hat{n}_{\sigma,i}$$
$$+ \sum_i \left(\frac{c_0}{\Delta}\hat{n}_{\uparrow,i}\hat{n}_{\downarrow,i} + \frac{c_\ell}{\Delta}\hat{n}_{\uparrow,i}\hat{n}_{\downarrow,i+s} + \frac{c_\ell}{\Delta}\hat{n}_{\uparrow,i+s}\hat{n}_{\downarrow,i}\right), \quad (65)$$

where $\mathcal{T} = \hbar^2/2m\Delta^2$, $\mathcal{U} = m\omega^2/2$, $\hat{n}_{\sigma,i} = \hat{a}^\dagger_{\sigma,i}\hat{a}_{\sigma,i}$ is the spin-$\sigma$ particle number operator at site $i$, and $s$ represents the distance $\ell$ in lattice sites, i.e., $\ell = s\Delta$.

## VI. DETAILS OF NUMERICAL SIMULATIONS

To study the many-body dynamics of the considered problem we employ matrix product states (MPS) [25, 26] ansatz with open boundary conditions where the system of size $L = 10$ has edges located at $x_{\text{OBC}} = \pm L/2$. The $N_s$ sites, representing the lattice in a discretized space, are located at positions $x_{j=1,\ldots,N_s} = -L/2 + \Delta(j - 1/2)$ with $\Delta = L/N_s$. The harmonic confinement is realized by employing the trap of frequency $\omega = \sqrt{2}$ centered at $x = 0$, which guarantees that particles in the systems under considerations ($M \leq 3$) never feel the hardwall boundaries. The initially separated state is prepared with the help of a standard density matrix renormalization group (DMRG) method [25–29] in the presence of a Gaussian barrier $V_b(x) = Ae^{-x^2/2\sigma^2}$, starting with a state where fermions of opposite spins occupy sites in different halves of the trap far from $x = 0$. For this purpose we used $A = 10(M + 1)$ and $\sigma = 0.2$.

Generally, time evolution within the MPS approach is error-prone and computationally demanding. It is due to the fact that together with a ballistic increase of an entanglement entropy in the course of dynamics, it is required to deal with increasing bond dimension.

This issue was partially overcome by a recent development of the time-dependent variational principle (TDVP) algorithm [30–32], which allows for a time propagation with a fixed bond dimension (one-site TDVP). It is much less error-prone in comparison with other earlier techniques like the time evolving block decimation (TEBD) routine [33]. Nevertheless, to reproduce a system dynamics properly, the bond dimension has to be first enlarged enough, which can be done for example as in the "hybrid" TDVP strategy, where the desired bond dimension is obtained through the initial application of the two-site TDVP scheme [33–35]. Here, we employ a recent approach in which the basis for MPS $|\Psi\rangle$ is enlarged by means of a subspace expansion by global Krylov vectors $\{|\Psi\rangle, \hat{H}|\Psi\rangle, \ldots, \hat{H}^{k-1}|\Psi\rangle\}$, where $\hat{H}$ is the Hamiltonian governing the system dynamics and $k$ denotes the so-called Krylov order. The method is meticulously described in Ref. [36], where the authors argue that this strategy is more accurate and more efficient than the two-site TDVP approach. Moreover, this method turns out to be more reliable in the case of beyond-nearest-neighbor interactions. We use the algorithm implementation based on the ITensor library provided by the authors of [36] under the ITensor/TDVP repository. The calculations

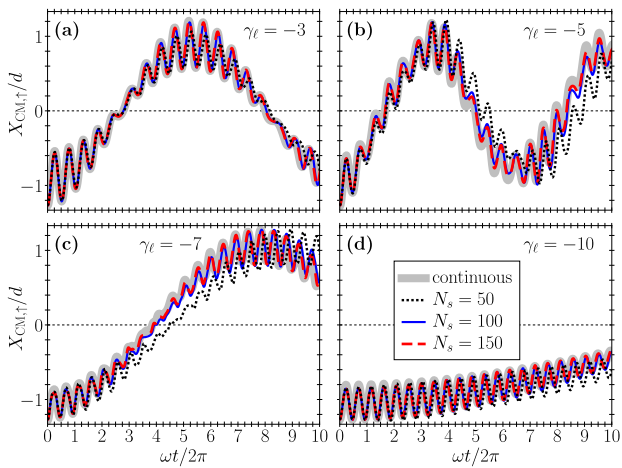

FIG. 6. Comparison between dynamics of the center-of-mass of the spin-$\uparrow$ component obtained in continuous and discretized models with $M = 1$, different number of sites $N_s = 50, 100, 150$, and different beyond contact attraction $\gamma_\ell = -3, -5, -7, -10$.

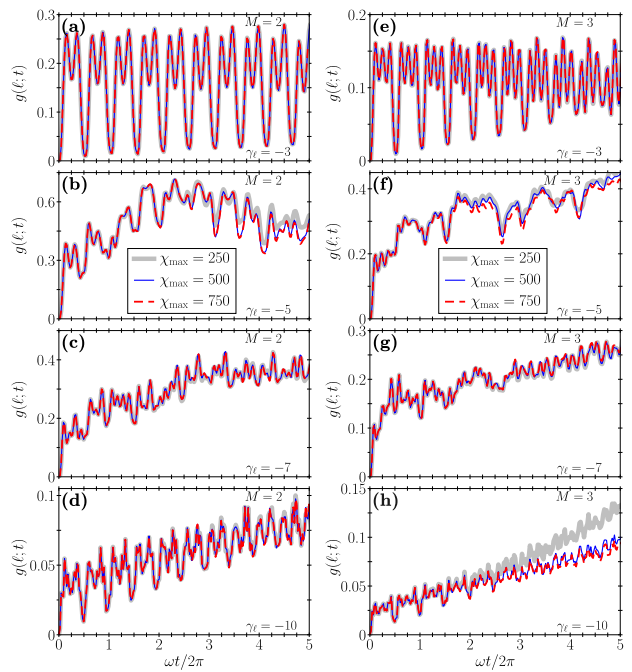

FIG. 7. Comparison between time evolution of $g(\ell; t)$, calculated for different beyond-contact attraction strengths $\gamma_\ell = -3, -5, -7, -10$, when setting different maximal bond dimension $\chi_{\max} = 250, 500, 750$. Left column, (a)–(d), corresponds to the results obtained for $M = 2$, while right column, (e)–(h), shows $g(\ell; t)$ computed in the $M = 3$ case.

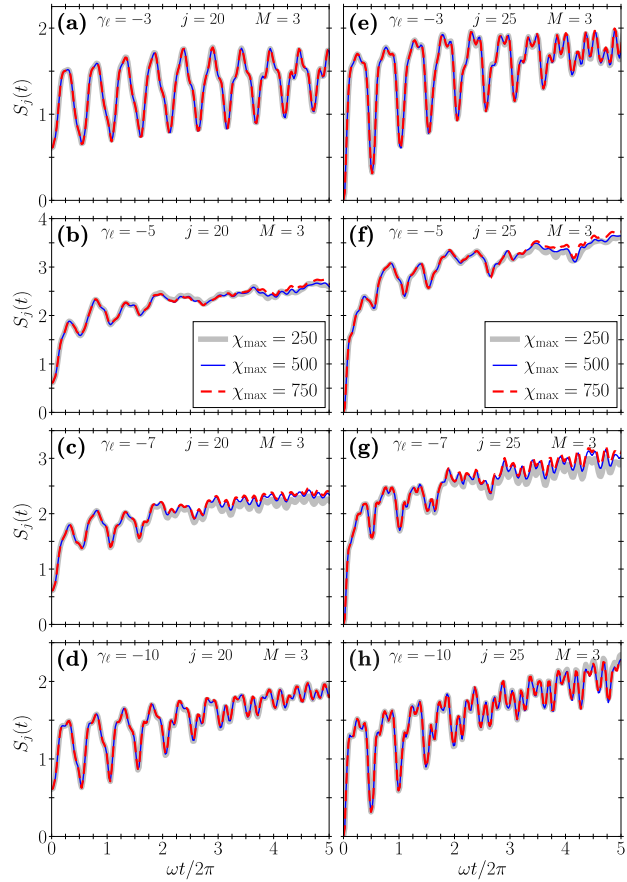

FIG. 8. Time evolution of the von Neumann entanglement entropy $S_j(t)$ calculated across the bond between $j$ and $j+1$ lattice sites for $M=3$, $N_s=50$, and two central bonds $j=20$ (left column) and $j=25$ (right column). The results obtained for different maximal bond dimensions $\chi_{\max}=250,500,750$ are very similar in the whole considered range of beyond contact attraction strength $\gamma_\ell = -3, -5, -7, -10$.

we performed with the time step $\Delta t = 0.02$ and $k=3$, where the truncation error of each application of $\hat{H}$ to $|\Psi\rangle$ and the truncation error controlling diagonalization of the sum of the reduced density matrices were chosen to be equal $10^{-8}$. Additionally, we preserve exact unitarity by imposing no truncation during one-site TDVP sweeps. This choice of parameters guarantees a balance between cost and accuracy, when testing in the most demanding $M=3$ case. We let the bond dimension $\chi$ of the time-evolved MPS grow up to pre-determined value $\chi_{\max} = 750$ and switch off the global basis expansion routine when $\chi_{\max}$ is reached.

First, we benchmark obtained results in the $M=1$ case, where the exact results in the continuous space are known. Namely, we consider the system with different numbers of sites $N_s = 50, 100, 150$, but with the same $\ell = 0.2$ in the three-delta $W(x)$ potential. Since $\Delta = L/N_s$, in order to compare the numerical results with the analytical predictions, it is required to take into account different beyond-contact attraction term in the discretized space, i.e. different values of the integer $s$ in the Hamiltonian (65). That is, while for $N_s = 50$ we employ nearest-neighbor attraction ($s=1$), for $N_s = 100$ and $N_s = 150$ we incorporate next-to-nearest-neighbour ($s=2$) and second-next-to-nearest-neighbour ($s=3$) interactions, respectively. The comparison between dynamics of the center-of-mass position of ↑-component, $X_{CM,\uparrow}$, determined analytically in the continuous model and numerically in the discretized system for $\gamma_\ell = -3, -5, -7, -10$ is illustrated in Fig. 6. Note that the results obtained for $N_s = 100$ and $N_s = 150$ are almost identical to the analytical ones, showing the convergence and reliability of the TDVP evolution we employed. On the other hand, in the case of $N_s = 50$ the considered center-of-mass position evolves slightly different in comparison with the analytical results. Nevertheless, the discreteized system behavior is still very similar to the continuous one, which allows us to suppose that conclusions we formulated basing on the analysis performed in the discretized model with $N_s = 50$ are also valid in the continuous systems.

Secondly, we show the convergence of our main results determined for $N_s = 50$ by comparing $g(\ell; t)$ obtained with different maximal bond dimension $\chi_{\max} = 250, 500, 750$ for $M=2$ and $M=3$ and different beyond-contact attraction strengths $\gamma_\ell = -3, -5, -7, -10$ (see Fig. 7). In addition, in Fig. 8 we present the comparison between dynamics of the von Neumann entanglement entropies, $S_j$, calculated across the bond between $j$ and $j+1$ sites for $M=3$, different $\chi_{\max}$, and two central bonds $j = 20, 25$.


\end{document}